\documentclass[aip,graphicx]{revtex4-1}
\usepackage[utf8]{inputenc}
\usepackage[T1]{fontenc}
\draft 
\usepackage{graphicx}
\usepackage{wrapfig}
\usepackage{textcomp}
\usepackage{subcaption}
\usepackage{braket}
\usepackage{bbold}
\usepackage{amsmath}
\usepackage{bm}
\usepackage{xcolor}
\usepackage{soul}
\usepackage{xspace}

\begin{document}


\title{Fool's gold: ligand-receptor interactions and the origins of life} 

\author{Betony Adams}
\affiliation{School of Data Science and Computational Thinking and Department of Physics, Stellenbosch University, South Africa}
\affiliation{National Institute for Theoretical and Computational Sciences, South Africa}
\affiliation{The Guy Foundation, Dorset, UK}
\author{Angela Illing}
\affiliation{Independent science illustrator}
\author{Francesco Petruccione}%
\affiliation{School of Data Science and Computational Thinking and Department of Physics, Stellenbosch University, South Africa}
\affiliation{National Institute for Theoretical and Computational Sciences, South Africa}
\date{\today}

\begin{abstract}
The origins of life is a question that continues to intrigue scientists across disciplines. One theory -- the iron-sulphur theory -- suggests that reactions essential to the synthesis of biological materials got their catalytic `spark' from mineral surfaces such as iron pyrite, commonly known as fool's gold. Additionally, the binding affinity of the ligands synthesised in this `surface metabolism' acted as an early version of natural selection: more strongly-binding ligands were accumulated into further autocatalytic reactions and the aggregation of complex biological materials. Ligand-receptor binding is thus fundamental to the origins of life. In this paper, we use the iron-sulphur theory as a lens through which to review ligand-receptor interactions as they are more commonly understood today. In particular we focus on the electron tunnelling theory of receptor activation that has emerged from research into quantum biology. We revisit criticism against this theory, particularly the lack of evidence for electron transfer in receptors, to see what insights might be offered by ligand-receptor interactions mediated by iron pyrite at the origins of life. What emerges from this comparison is the central importance of redox activity in receptors, in particular with respect to the recurring presence of the disulphide bond. While the paper is a speculative exercise, we conclude that conductivity in biomolecules, particularly the selective conductivity conferred by appropriate ligand-receptor binding, is a powerful tool for understanding diverse phenomena such as pharmacological potency and viral infection. As such it deserves further investigation. 

    \end{abstract}
\maketitle 

\begin{figure}[hbt!]
\centering
\includegraphics[scale=0.27]{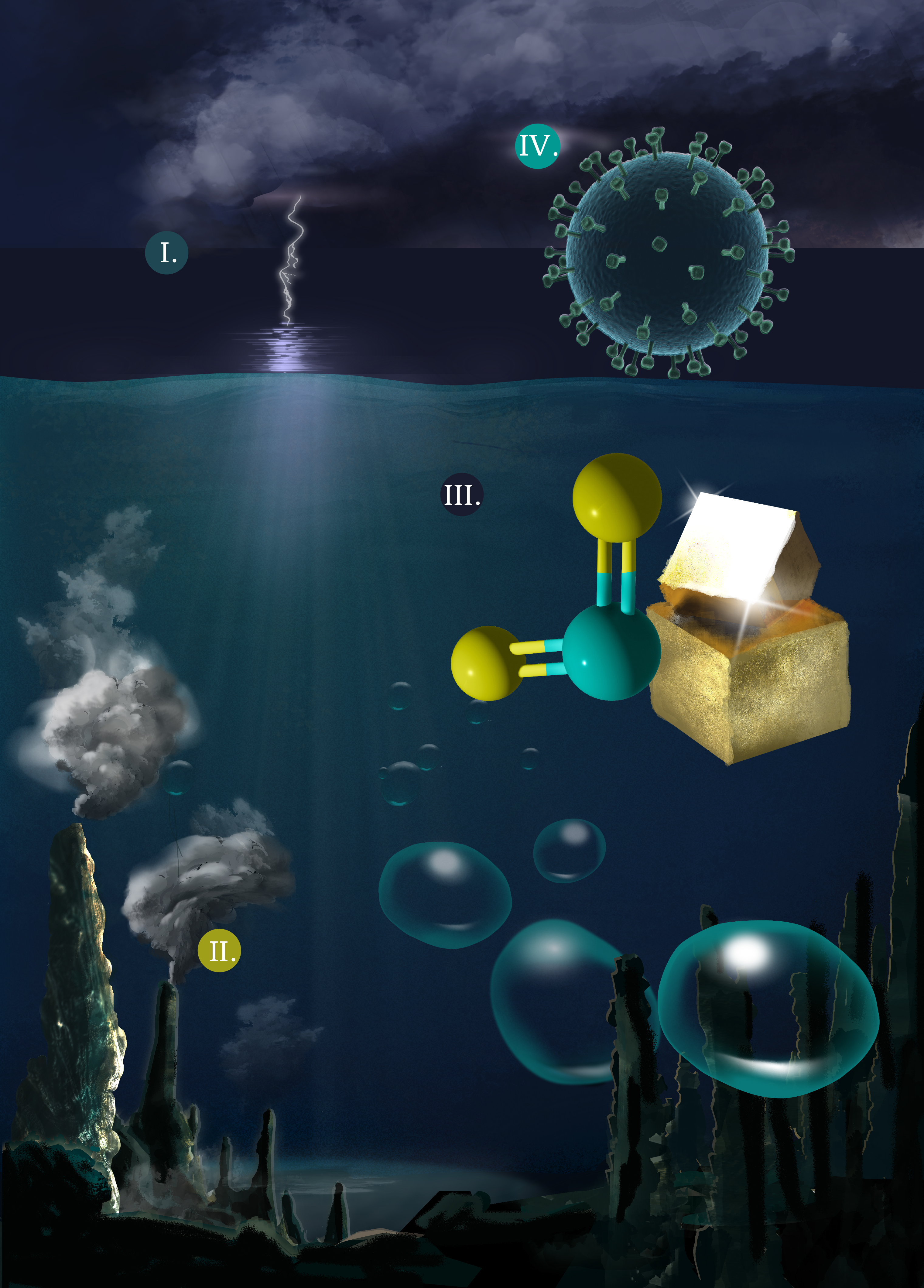}
\caption{Electron transfer, ligand-receptor interactions and the origins of life. I. The originary spark. II. Fool's gold, metal catalysts at the origins of life. III. Ligand-receptor activation and quantum biology. IV. Conductivity as a measure of binding affinity.}
\end{figure}
\section{The originary spark}
\begin{wrapfigure}{R}{0.45\textwidth}
\centering
\includegraphics[width=0.4\textwidth]{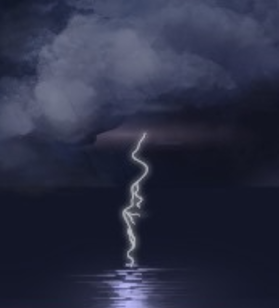}
\end{wrapfigure}
An accurate definition of `life' is a question that has preoccupied thinkers across disciplines, from philosophy to physics. While there is yet to be any consensus on this definition it is generally agreed that living systems are distinguished by their unique materials. Perhaps most intriguing of these are proteins. Proteins, and their constituent amino acid molecules, have been recognised as distinct materials for centuries \cite{reynolds}. And, while it remains impossible to synthesise life from scratch, organic molecules such as amino acids have been synthesised from inorganic molecules in a number of different experiments \cite{miller,ring,jiang}. \\
\\
Most famous of these is the Miller-Urey experiment, the results of which were published in 1953 \cite{miller}. Hydrogen, ammonia and methane were combined in a beaker to mimic the Earth's prebiotic ocean and atmosphere, while an electrical spark was passed through the mixture to simulate lightning. The experiment succeeded in synthesising a number of the amino acids integral to biology \cite{miller}. While this simple spark was sufficient for the first steps in the chemistry of life, it does not explain how amino acids form more complex biological molecules such as enzyme proteins, which are responsible for many biological processes.\\
\\
Indeed, one of the arguments against metabolism-first approaches in origins of life research, is the lack of enzymes necessary to catalyse metabolic processes, such as biosynthesis \cite{davey,anet}. Scientists have thus turned to alternative theories as to what original catalytic `spark' drove the redox reactions so central to biology. One of these theories suggests that life first formed on the surfaces of minerals containing iron sulphur compounds, such as pyrite or fool's gold.
\newpage
\section{Fool's gold, metal catalysts at the origins of life}
\begin{wrapfigure}{L}{0.35\textwidth}
\centering
\includegraphics[width=0.3\textwidth]{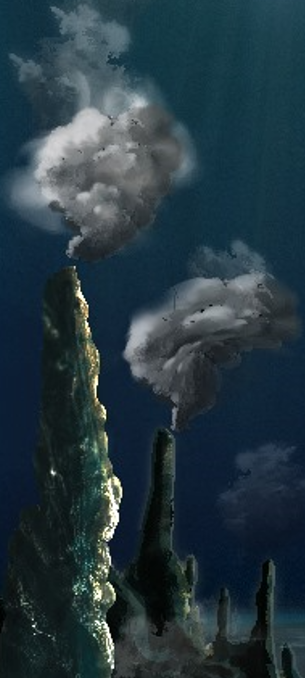}
\end{wrapfigure}
The iron sulphur theory of life's origins is initially attributed to G\"{u}nter W\"{a}chtersh\"{a}user who, in a 1988 paper, addressed the question of an originary spark \cite{wach1}. 
\begin{quote}
    Any theory that might attempt to explain the autotrophic origin of life is faced with the unsolved problem of 
    the aboriginal energy flow, the prime mover of all subsequent processes of evolution. Two basic types of chemical energy may be considered: photochemical energy (photo-autotrophy) and inorganic redox energy (chemo-lithoautotrophy) \cite{wach1}.
\end{quote}
He discounts photochemistry, arguing that the necessary light-reactive cellular apparatus (chlorophylls and retinals) were not yet available. Instead, he suggests that the exergonic formation of pyrite could have kick-started the reactions necessary to life's origins, potentially in oceanic hydrothermal vents \cite{wach1}. In another paper, later in the same year, he goes on to outline the role that pyrite might play in surface metabolism at the very earliest stages of the evolution of living systems \cite{wach2}. He describes an early version of natural selection in which the first organic molecules are synthesised on mineral surfaces, such as pyrite, and then selected for by aggregation, depending on how strongly they adhere to this surface. These bound molecules then act as ligands in further reactions, forming ever more complex biological molecules \cite{wach2}.\\
\\
The iron sulphur world theory is not universally accepted \cite{lane}. However, it does seem likely that, given the lack of complex proteins to act as catalysts, metal ions and metal complexes may have played a pivotal role in driving reactions at the origins of life, being subsequently incorporated into the materials out of which living organisms are made \cite{belmonte,rossetto,moran}. Transition metals are redox active, being electron donors, a role that is reflected in the numerous transition metal centres that facilitate electron transfer reactions at the active sites of biological molecules such as oxireductases \cite{mele,kim}. Indeed, structural similarities between metal-ligand-binding sites in modern proteins suggest that metal-facilitated electron transfer played an essential role in the origins of life, giving rise to the electron transfer chains that are of fundamental importance to life as we recognise it today \cite{rossetto,raanan,bromberg}.\\
\\
In the context of this paper, which is preoccupied with the mechanisms by which biological receptors work, the iron sulphur theory of biological origins may offer some insights. In the broadest sense, all biological activities involving interactions between constituent reactants might be thought of as involving receptors, with the more complex reactant labelled the receptor (enzyme, immunoglobulin, synapse) as opposed to the simpler ligand (substrate, antigen, neurotransmitter) \cite{klotz}. \\
\\
The iron sulphur hypothesis of surface metabolism suggests that binding affinity played an important selective role in the evolution of life, and that the binding affinity of these originary ligands further catalysed biological reactions such as biosynthesis \cite{wach2}. Receptor binding, in this sense, is thus central to the origins of life. The implication of binding affinity in the aggregation of biological materials also offers potential ways to explain some of the open questions in biology, such as the chiral asymmetry of biological materials. Chirality matters in biology. It is well established that ligands with different chiralities can have different binding affinities \cite{terauchi,schneider}. If binding affinity was implicated in prebiotic selection of ligands at the origins of life, then this could result in the `survival' of the specific chiral biases we see today.\\
\\
In addition to this, the involvement of metal ions suggests that electron transfer was also important. Ligand-receptor interactions at the origins of life would thus prioritise binding affinity and redox potential. It is interesting to consider how these design principles, and the fossil record of these original molecules, may have shaped receptor binding and activation, as we more commonly understand it today. On a very simplistic level, receptor binding and activation is often described as using a `lock-and-key' mechanism, where the shape of the ligand is the deciding factor as to which receptor it will activate \cite{tripathi}. However, research in quantum biology has suggested that electron tunnelling may actually be involved and that ligands act less like puzzle pieces than plugs, with binding affinity related to electron transfer \cite{turin96,horsfield}.
\newpage

\section{Ligand-receptor activation and quantum biology}
\begin{wrapfigure}{R}{0.45\textwidth}
\centering
\includegraphics[width=0.4\textwidth]{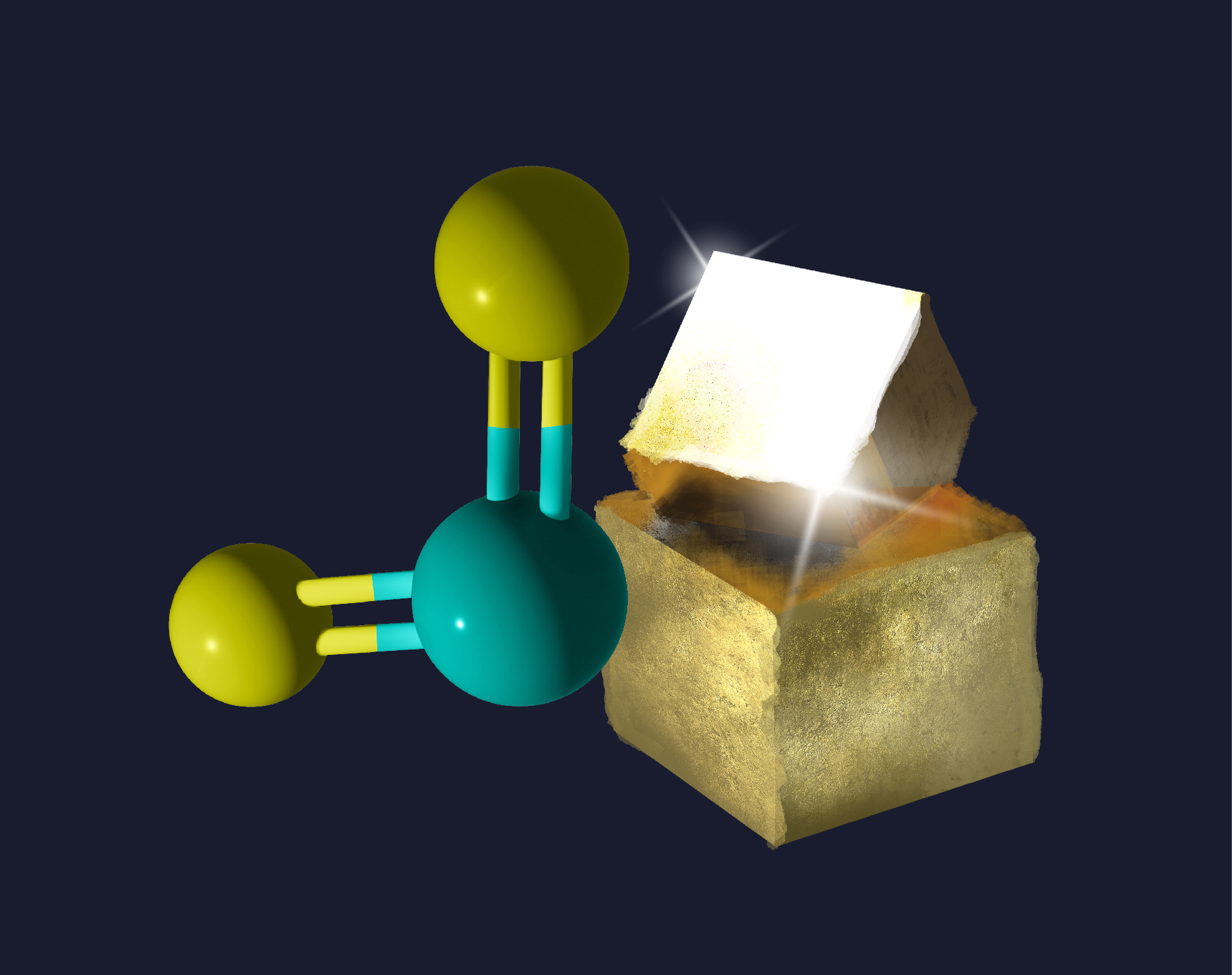}
\end{wrapfigure}
Quantum biology is the study of quantum effects in biological systems, particularly how these effects are related to biological function. The transfer of electrons is a fundamental topic in quantum biology, in a number of different contexts, including receptor activation. While we have defined receptors very generally in the last section, quantum biology research focuses more specifically on transmembrane receptors such as G-protein coupled receptors (GPCRs), particularly olfactory receptors \cite{dyson,turin96,horsfield} and to a lesser degree neuroreceptors \cite{hoehn,chee2}. The vibration-assisted tunnelling theory of receptor activation suggests that the vibrational spectrum of the ligand facilitates electron tunnelling in the appropriate receptor \cite{turin96,horsfield}.\\
\\
The theory has met with some scepticism. One of the often-cited arguments against the theory is the lack of good evidence for electron transfer in receptors. Recent research, from proponents of the vibration tunnelling hypothesis, has identified some possible candidate sites that counter this caveat \cite{horsfield,gehrckens,gehrckens2}. The requirements for electron transfer in a receptor include a donor molecule with a highest occupied molecular orbital (HOMO) that is higher in energy than than the lowest unoccupied molecular orbital (LUMO) of its associated acceptor molecule \cite{horsfield}. One way in which this can be achieved is if a metal ion, which can shift a LUMO down in energy, is bound to the receptor protein \cite{horsfield, gehrckens}. In further research, a number of possible molecular wires, such as tryptophan molecules, are identified in GPCRs to support electron transfer \cite{gehrckens,gehrckens2}. Chromophores, which are light-reactive molecules such as tryptophan, are abundant in biological materials, and are central to much of the electronic charge transfer research in quantum biology \cite{adams}.\\
\\
Interestingly, the end point of this electron transfer, regardless of path, is suggested to be the highly conserved disulphide bond that is integral to receptor function \cite{gehrckens2}. If pyrite -- iron disulphide -- were pivotal in surface metabolism and aggregation through binding affinity at the origins of life, it might not be surprising to find structural similarities -- such as the disulphide bond -- at play in the binding affinity of receptors in living organisms today. The disulphide bond is central to protein stability, but there is growing evidence that it can also act as a `redox switch', the function of which is implicated in signalling processes \cite{west}. Just how this works in transmembrane receptors, such as GPCRs, is a largely unexplored phenomenon.\\
\\
While the focus in quantum biology has been predominantly on GPC receptors, disulphide bonds have been implicated in the activity of other receptors as well. In a previous paper we investigated electron tunnelling in the spike protein-ACE receptor interaction by which the SARS-CoV-2 virus infects host cells \cite{adams2}. Evidence suggests that this interaction might involve redox activity \cite{singh,keber,hati}. Interestingly, the infectivity of SARS-CoV-2 may depend on the redox potential of a disulphide bond, with disease resistance conferred by the lack of a redox-active disulphide \cite{singh,keber}. Reduction of the disulphide bonds of ACE2 and SARS-CoV-2 spike proteins also impairs binding affinity \cite{hati}. \\
\\
Vibration assisted tunnelling in receptors is still largely theoretical. However, given the importance of charge transfer in biology, and the possibility that receptors have structural features that support charge transfer, the electronic properties of ligand-receptor complexes should be a subject of interest. A better understanding of these properties would have a number of important implications, not least of which is how viruses or drugs hijack host cell receptors.
\newpage
\section{Conductivity as a measure of binding affinity}
\begin{wrapfigure}{L}{0.45\textwidth}
\centering
\includegraphics[width=0.4\textwidth]{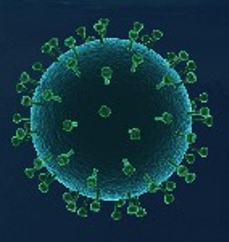}
\end{wrapfigure}
Proteins are conventionally considered to be inefficient conductors. Revisiting this assumption from the point of view of receptor-mediated electron transfer, conductivity offers a novel and useful way to classify binding affinity. Comparisons of protein conductivity using ligand specific and non-specific binding to scanning tunnelling microscope (STM) contacts, demonstrates highest conductivity for the case in which the binding was ligand specific \cite{zhang}. Ligand-receptor binding affinity appears to facilitate protein conductivity. This method was used for accurate single molecule identification in the context of viral infection \cite{zhang}. The experiment used appropriate peptide epitopes tethered to the contacts of an STM to determine whether Ebola-specific antibodies were present in a solution or not. The presence of antibodies specific to the contact epitopes resulted in a current, whereas non-specific proteins generated no signal \cite{zhang}.\\
\\
The authors of this study go on to discuss the possible future applications of this technique, which include sensitive background-free single-molecule detection as well as the investigation of protein structure and motion \cite{zhang}. We add to this the possibility of using conductivity as a measure of binding affinity or ligand efficacy and potency. It would be highly instructive to test whether these ligand properties in some way correlate with receptor conductivity. Efficacy and potency are important ligand properties but the exact mechanisms underlying these properties are still not properly understood \cite{babu}. \\
\\
In addition to pharmacological insights, the correlation of binding affinity with receptor conductivity might be used to investigate the relative infectiousness of viral mutants such as the different SARS-CoV-2 strains. The SARS-CoV-2 spike protein binds to and invades host cells using ACE2 receptors \cite{tang,ni}. Mutations in this spike protein led to fears of increased viral increased transmissability \cite{volz,zhang2}. It could be interesting to compare the conductivity of ACE2 receptors for STM contacts functionalised with different spike protein mutants.\\
\\
While this paper is a speculative exercise, we feel there is sufficiently intriguing evidence to support increased research into the electronic properties of ligand-receptor interactions. To summarise: properties of ligand-receptor interactions such as binding affinity and electron transfer may have played a pivotal part in the origins of life. Binding affinity, in theories of pyrite-mediated surface metabolism, acted as an early form of molecular natural selection through bonding strength, in particular charge-mediated anionic bonding \cite{wach2}. Binding affinity was also implicated in autocatalytic activity, with metals potentially playing a role in early electron transfer reactions \cite{rossetto,raanan,bromberg}. \\
\\
Binding affinity and electron transfer thus seem fundamental to ligand-receptor interactions at the origins of life. A better understanding of these design principles may have relevance to our understanding of receptor binding in the more specific contexts we currently study. The electronic properties of proteins demonstrate nature's flexible engineering. While proteins have been conventionally thought of as inefficient conductors the reality may be infinitely more interesting. It has been suggested that many biomolecules actually sit at the critical quantum state between metal and insulator \citep{vattay}. Ligand-receptor interactions might thus be more fruitfully thought of as the switch that acts on this criticality, conferring selective conductivity through binding affinity.

\bibliography{name}

\end{document}